\documentclass{iau}
\usepackage{graphicx}
\usepackage{subfigure}% 
\title[Effects of primordial magnetic fields on CMB] %% give here short title %%
{Effects of primordial magnetic fields on CMB}

\author[H\'ector J. Hort\'ua \& Leonardo Caste\~neda]   %% give here short author list %%
{H\'ector J. Hort\'ua$^{1}$
 \and Leonardo Casta\~neda$^1$}
\affiliation{$^1$Grupo de Gravitaci\'on y Cosmolog\'ia, Observatorio Astron\'omico Nacional, \\ Universidad Nacional de Colombia, Cra 45 \# 26-85, Bogot\'a D.C.,  Colombia  \\email: {\tt hjhortuao@unal.edu.co}  \\[\affilskip]
}
\pubyear{2008}
\volume{xxx}  %% insert here IAU Symposium No.
\pagerange{119--126}
\jname{Title of your IAU Symposium}
\editors{A.C. Editor, B.D. Editor \& C.E. Editor, eds.}
\begin{document}

\maketitle

\begin{abstract}
The origin of large-scale magnetic fields is an unsolved problem in  cosmology. In order to overcome, a possible scenario comes from the idea that these fields emerged from a small primordial magnetic field (PMF), produced in the early universe. This field could lead to the observed large-scales magnetic fields   but also,  would have left an imprint on the cosmic microwave background (CMB). In this work  we  summarize some statistical properties of this PMFs  on the FLRW background. Then, we show the resulting PMF  power spectrum  using cosmological  perturbation theory  and  some  effects  of PMFs on  the  CMB anisotropies.
%% add here a maximum of 10 keywords, to be taken form the file <Keywords.txt>
\end{abstract}
\begin{figure}[h]
\centering
\subfigure[Magnetic density of PMF power spectrum  for different values of amplitude and spectral indices.]{%
\includegraphics[width=2.5in]{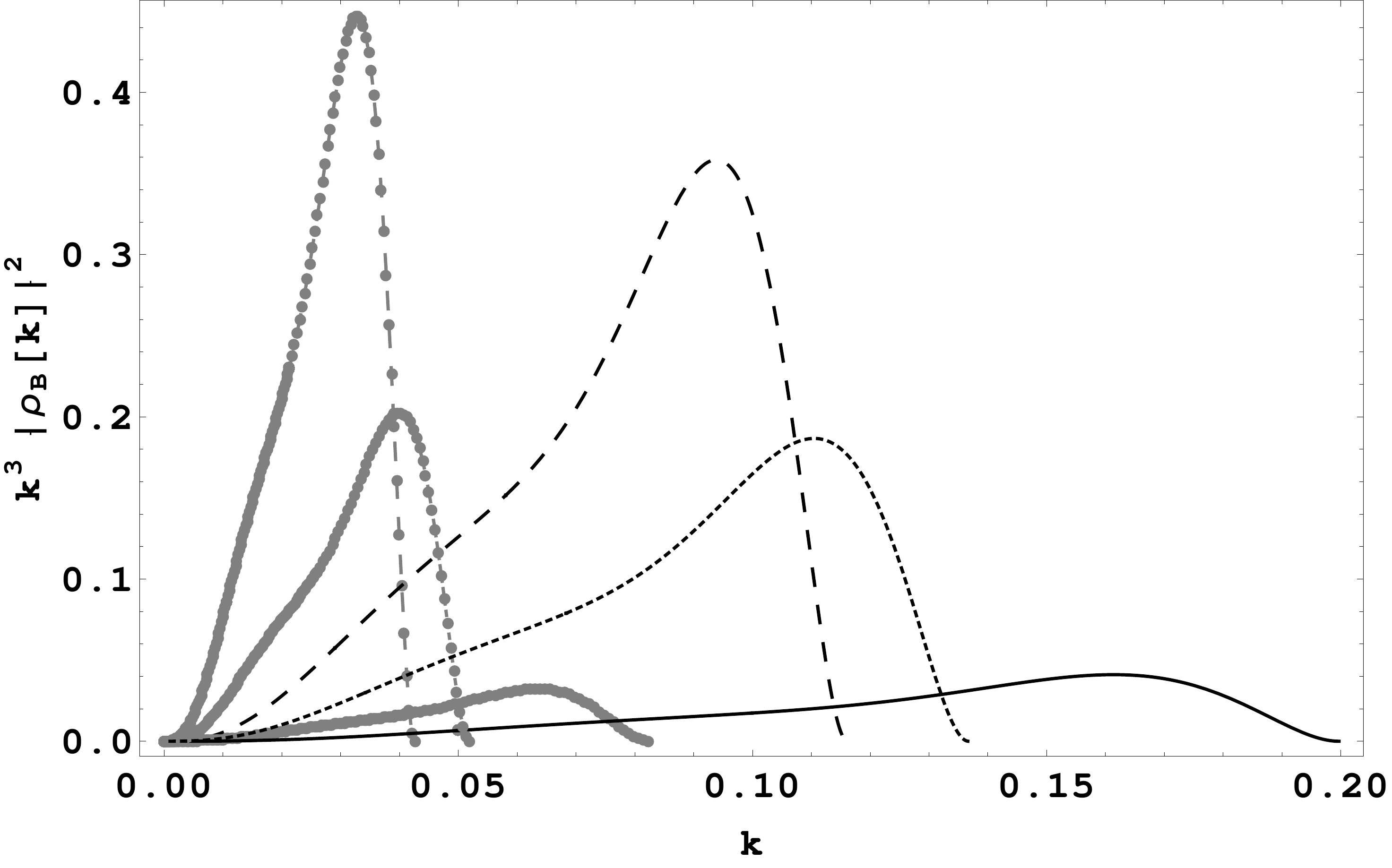}
\label{densidad1}}
\quad
\subfigure[Lorentz force spectra (black lines) and scalar anisotropic trace-free part (gray lines with points)   for  different spectral indices.]{%
\includegraphics[width=2.5in]{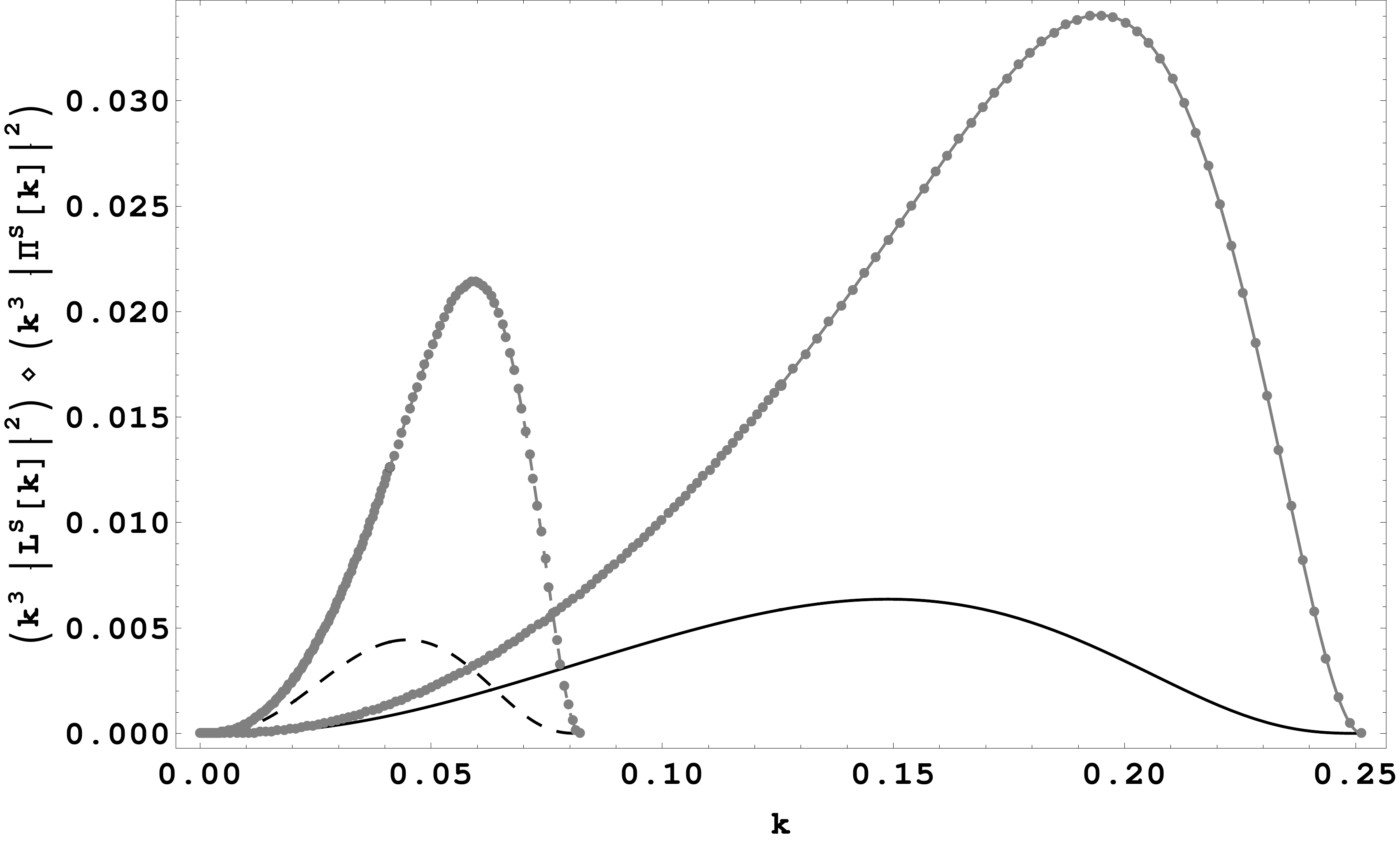}
\label{lorentz1}}
\subfigure[Cross-correlation of $|\rho_B\Pi^{(S)}|$ in gray with points and   $|\rho_BL^{(S)}|$ in black, 
for different 
 spectral indices.]{%
\includegraphics[width=2.5in]{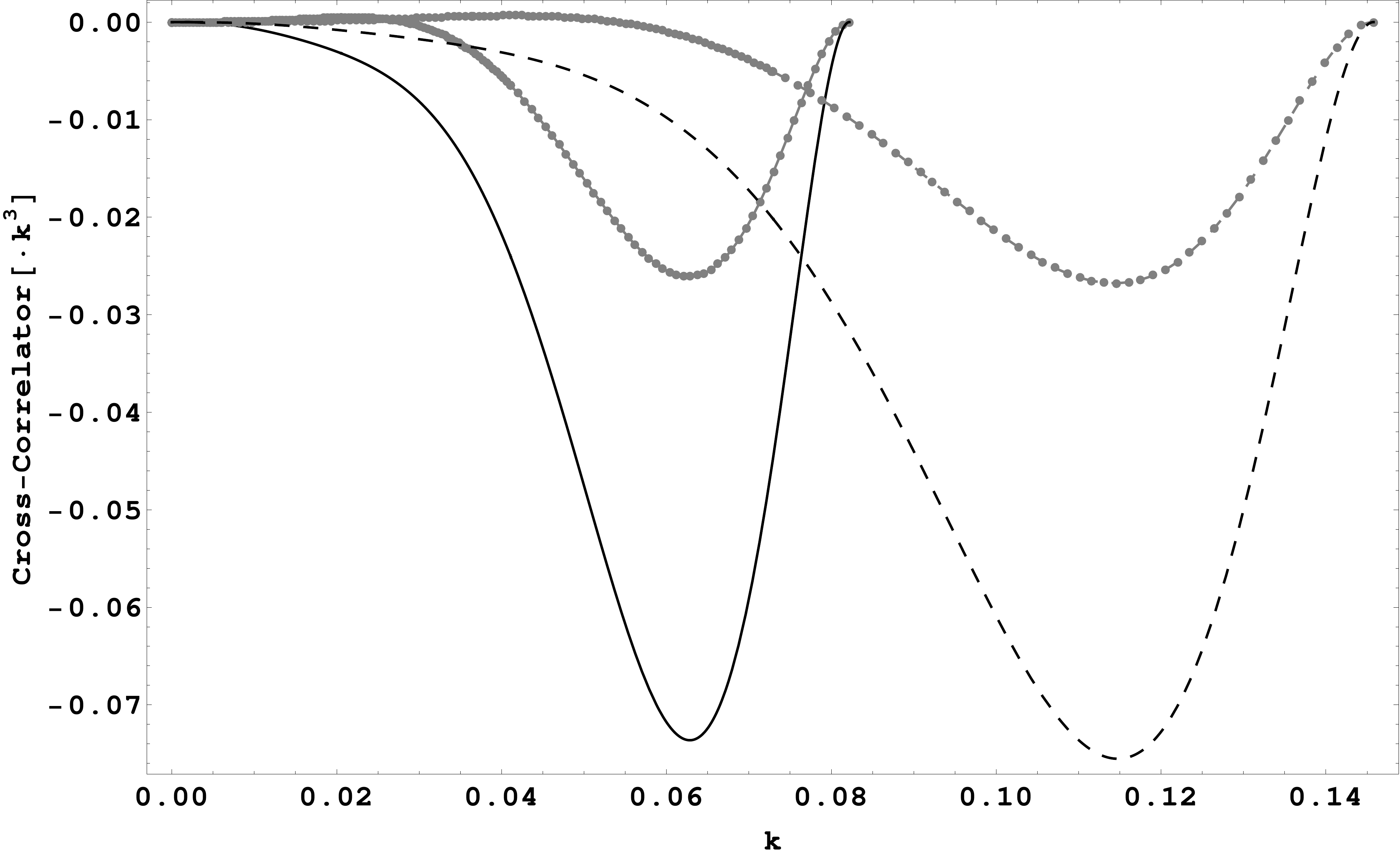}
\label{correlators}}
\quad
\subfigure[CMB temperature anisotropy
angular power spectrum  induced by scalar
magnetic mode. 
 ]{%
\includegraphics[width=2.5in]{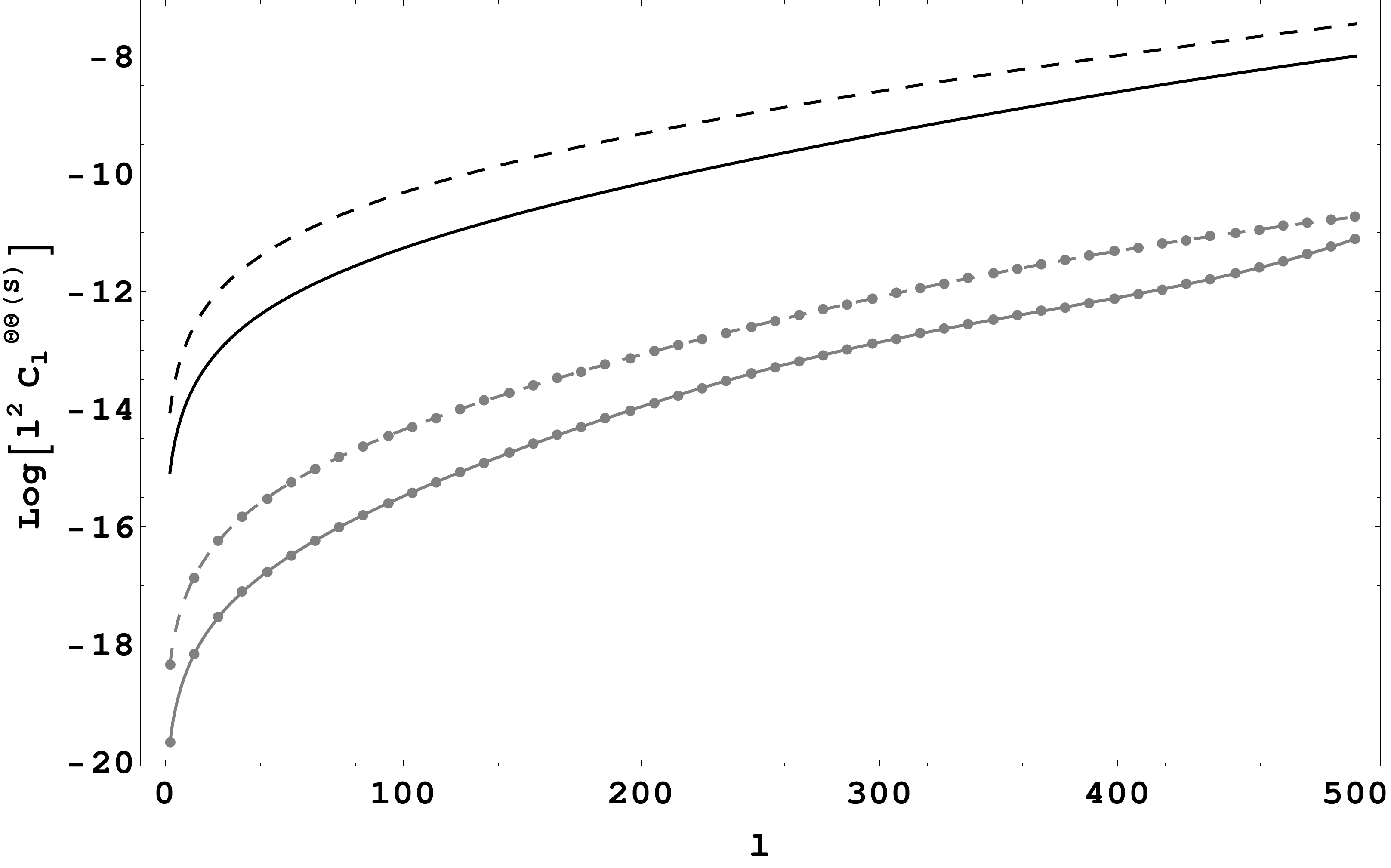}
\label{cL1}}
\caption{Correlators and power spectrum for CMB anisotropies due to a PMF.}
\label{fig:figure}
\end{figure}

Magnetic fields have been observed in all scales of the universe, from planets and stars to galaxies and galaxy clusters.
 However, the origin of such a magnetic field is still unknown. Some theories argue  magnetic fields we observe today has a primordial origin, indeed, there are some
processes in early epoch of the universe that would have created a small primordial magnetic field (PMF). 
%The origin of this PMFs can be searched as electroweak and QCD phase transitions, inflation, among others. 
If PMFs really were present, 
these could have some effect on Nucleosynthesis and would leave imprints in the   CMB  fluctuation.\\
{\underline{\it The model and  statistics for a stochastic PMF}}. We consider a causal stochastic PMF  generated after  inflation, thus the maximum coherence lenght for
the fields must be not less than  the Hubble horizon (\cite[ Kahniashvili et al. 2007]{tina}).  
Now, the PMF power spectrum  which is defined as the Fourier transform of the two point correlation  can be written as
\begin{equation}\label{PMFespectro}
\langle B_i^*(\textbf{k})B_j(\textbf{k}^\prime)\rangle =(2\pi)^3\delta^3(\textbf{k}-\textbf{k}^\prime)P_{ij}P_B(\textbf{k}),
\end{equation}
where $P_{ij}=\delta_{ij}-\frac{\textbf{k}_i\textbf{k}_j}{k^2}$ is a projector onto the transverse plane,  $P_B(\textbf{k})$ is the
PMF power spectrum. We focus our attention to the evolution of a causally-generated  PMF  parametrized by a power law with index $n \geq 2$,
 with  an ultraviolet cut-off $k_{D}$ and the dependence of an infrared cutoff, $k_m$. 
The power spectrum can be defined as
\begin{equation}\label{powerPMF}
P_B(k)=Ak^n \quad  \mbox{ with} \quad A=\frac{B^2_{\lambda}2\pi^2\lambda^{n+3}}{\Gamma(\frac{n+3}{2})}; \quad  \mbox{for  $k_{m} \leq k \leq k_{D}$},
\end{equation}
where $B_\lambda$ is the comoving PMF strength smoothing over a Gaussian sphere of comoving radius $\lambda$. The  energy density of PMF  and anisotropic trace-free part are  written as 
\begin{equation}\label{convdensity1}
\rho_B(k)=\frac{1}{8\pi}\int \frac{d^3k^\prime}{(2\pi)^3}B_l(k)B^l(|\textbf{k}-\textbf{k}^\prime|),
\end{equation}
\begin{equation}\label{convdensity}
\Pi_{ij}(k)=\int \frac{d^3k^\prime}{2(2\pi)^4}\left[B_i(k^\prime)B_j(|\textbf{k}-\textbf{k}^\prime|)-\frac{\delta_{ij}}{3}B_l(k^\prime)B^l(|\textbf{k}-\textbf{k}^\prime|) \right],
\end{equation}
(in the Fourier space) and we can write the Lorentz force on matter as $ \Pi^{(S)}=L^{(S)}(\textbf{x},\tau_0)+\frac{1}{3}\rho_B(\textbf{x},\tau_0)$, here $k=|\textbf{k}|$.   Now, we define the  two point correlation function as
\begin{equation}\label{corr}
\langle \Xi(k,\tau)\Xi^{*}(k^\prime,\tau)\rangle=(2\pi)^3 \left|\Xi(k,\tau)\right|^2\delta^3(\textbf{k}-\textbf{k}^\prime),
\end{equation}
where $\Xi(k,\tau)=\{\rho_B(k,\tau),  \Pi(k,\tau),  L(k,\tau)\}$,  and the cross correlation between them. Now, to calculate the power spectrum, we use the eqs.  (\ref{PMFespectro}), (\ref{convdensity1}),(\ref{convdensity}), (\ref{corr}) and  the  Wick's theorem assuming Gaussian statistics to evaluate the 4-point correlator of the PMF.  The power spectrum for  $\rho_B(k,\tau)$,  $\Pi(k,\tau)$,  $L_B(k,\tau)$ are given by
\begin{equation}
\left|\rho_B(k)\right|^2=\frac{1}{256\pi^5}\int d^3k^\prime(1+\mu^2)P_B(k^\prime)P_B(\left|\textbf{k}-\textbf{k}^\prime\right|),
\end{equation}
\begin{equation}
\left|L^{(S)}(k)\right|^2=\frac{1}{256\pi^5}\int d^3k^\prime[4(\gamma^2\beta^2-\gamma\mu\beta)
+1+\mu^2]P_B(k^\prime)P_B(\left|\textbf{k}-\textbf{k}^\prime\right|),
\end{equation}
\begin{equation}
\left|\Pi^{(s)}(k)\right|^2=\frac{1}{576\pi^5}\int d^3k^\prime[4-3(\beta^2+\gamma^2)+\mu^2
+9\gamma^2\beta^2-6\mu\beta\gamma]P_B(k^\prime)P_B(\left|\textbf{k}-\textbf{k}^\prime\right|),
\end{equation}
and for the scalar cross-correlation we have the relations
\begin{equation}
\left|\rho_B(k)L^{(S)}(k)\right|=\frac{1}{256\pi^5}\int d^3k^\prime[1-2(\gamma^2+\beta^2)
+2\gamma\mu\beta-\mu^2]P_B(k^\prime)P_B(\left|\textbf{k}-\textbf{k}^\prime\right|),
\end{equation}
\begin{equation}
\left|\rho_B(k)\Pi^{(S)}(k)\right|=\frac{1}{128\pi^5}\int d^3k^\prime\bigg(\frac{2}{3}-(\gamma^2+\beta^2)
+\mu\gamma\beta-\frac{1}{3}\mu^2 \bigg) P_B(k^\prime)P_B(\left|\textbf{k}-\textbf{k}^\prime\right|),
\end{equation}
 where $\beta=\frac{\textbf{k}\cdot(\textbf{k}-\textbf{k}^\prime)}{k\left|\textbf{k}-\textbf{k}^\prime\right|}, \, \mu=\frac{\textbf{k}^\prime\cdot(\textbf{k}-\textbf{k}^\prime)}{k^\prime\left|\textbf{k}-\textbf{k}^\prime\right|}, \,  \gamma=\frac{\textbf{k}\cdot\textbf{k}^\prime}{kk^\prime}$. Our results are in agreement with  (\cite[ Kahniashvili et al. 2007]{tina}) and (\cite[ Finelli et al. 2008]{pao}).\\
{\underline{\it The cutoff dependence and the concern scale}}. We work with a upper cutoff $k_D$ corresponds to the damping scale due to  magnetic  energy  dissipation into heat through the damping of the Alfv\'en  waves. The upper cutoff of PMF is given by equation (31) in  (\cite[ Hortua et al. 2014]{hector}).  The most general  scenario at PMFs  takes into  account a infrared cutoff $k_m$  for low values
of $k$ and   depends on the generation model of PMF (\cite[ Yamasaki, 2014]{yama}). We approximate this infrared cut-off as $k_m= \alpha k_D$ with $0<\alpha<1$.\\
{\underline{\it PMF power spectra}}. In the figure \ref{densidad1}, we show the magnetic energy density convolution and its dependence with the spectral index ($n=7/2$ for black and $n=2$ for gray with points) and the amplitude of PMF at a scale of $\lambda = 1$Mpc (thick for $B=1nG$, large dashed for $10nG$ and  small dashed for $5nG$). In the figure \ref{lorentz1},  the  Lorentz force and the anisotropic part  spectra are shown with $n=2$ for dashed lines and $n=4$ for continue lines  and  fixed  an  amplitude of $B^2_\lambda=1nG$ at a scale of $\lambda = 1$Mpc.  Figure \ref{correlators}  shows  the cross-correlation between the energy density with  Lorentz force and the anisotropic trace-free part ($n=2$ continue lines and $n=3$ for dashed lines).  We notice that the cross correlation is negative in all range of scales for Lorentz force,  while to be  negative for values of $k \geq 0.05$ (with $n=3$) and $k \geq 0.03$ (with $n=2$) for anistropic trace-free.  The integration domain and the effects of the upper and lower cutoff are described in (\cite[ Hortua et al. 2014]{hector}). \\
{\underline{\it CMB angular power spectra}}. If  PMFs really were present before the recombination era, these     would leave imprints on CMB. Using the total angular momentum formalism, the scalar angular power spectrum of the CMB temperature  anisotropy  is given by
\begin{equation}\label{momenttemp}
(2l+1)^2C_l^{\Theta \, \Theta}=\frac{2}{\pi}\int dkk^2 \Theta_l^{(S)\, *}(\tau_0,k)\Theta_l^{(S)}(\tau_0,k),
\end{equation}
where  $\Theta_l^{(S)}(\tau_0,k)$ are the temperature fluctuation multipolar moments.  Now, considering only the fluctuation via PMF perturbation, (\cite[ Kahniashvili et al. 2007]{tina})  found that   temperature anisotropy multipole moment  becomes $\frac{\Theta_l^{(S)}(\tau_0,k)}{2l+1}\approx \frac{-8\pi G}{3k^2 a_{dec}^2}\rho_B(\tau_0,k)j_l(k\tau_0)$,
where $a_{dec}$ is the  scalar factor at decoupling, $G$ is the Gravitational constant and  $j_l$ is the spherical Bessel function.  Substituting the last expression  in equation (\ref{momenttemp}), the CMB temperature anisotropy angular power spectrum is given by
\begin{equation}\label{eqcls}
l^2C_l^{\Theta \, \Theta\,(S)}=\frac{2}{\pi}\left(\frac{8\pi G}{3a_{dec}^2}\right)^2\int_0^\infty\frac{\left|\rho_B(\tau_0,k)\right|^2}{k^2}j_l^2(k\tau_0)l^2dk,
\end{equation}
where for our case, the integration  is  only up to  $2k_D$, since it is the range where energy density power spectrum is di\-fferent from zero.
  The result of the angular power spectrum  induced by scalar magnetic perturbations given by eq. (\ref{eqcls}) is shown in the figure \ref{cL1}.  
Here, the black thick line defines the power spectrum for $n=2$ and the black dashed line  for   $n=5/2$, both with a strength of $B=1$nG. The gray thick line with points is for $n=2$ and the other one is for $n=5/2$, the gray lines with points are for $B=10$nG.\\
{\underline{\it Acknowledgments}}. H\'ector J. Hort\'ua acknowledges the IAU-S306 Grant.

\end{document}